\documentclass[preprint,prb,groupedaddress,preprintnumbers,amsmath,amssymb,floatfix]{revtex4-1}
\usepackage{graphicx}
\usepackage{amssymb}
\usepackage{dcolumn}
\usepackage{bm}
\begin{document}

\title{Annealing of single lamella nanoparticles of polyethylene}

\author{Christophe N. Rochette$^1$, 
Sabine Rosenfeldt$^1$,
Katja Henzler$^2$,
Frank Polzer$^2$, 
Matthias Ballauff$^{2,*}$,
Qiong Tong$^4$, 
Stefan Mecking$^4$,
Markus Drechsler$^5$,
Theyencheri Narayanan$^6$
Ludger Harnau$^{3,*}$}

\affiliation{$^1$Physikalische Chemie I, University of Bayreuth, 
95440 Bayreuth, Germany\\
$^2$Soft Matter and Functional Materials, Helmholtz-Zentrum Berlin, 
14109 Berlin, Germany\\
$^3$Max-Planck-Institut f\"ur Intelligente Systeme, Heisenbergstr.~3, 
70569 Stuttgart, Germany,
\\and Institut f\"ur Theoretische und Angewandte Physik, Universit\"at Stuttgart,
Pfaffenwaldring 57, 
70569 Stuttgart, Germany\\
$^4$Department of Chemistry, University of Konstanz, Universit\"atsstrasse 10, 
78457 Konstanz, Germany\\
$^5$Makromolekulare Chemie II, University of Bayreuth, 
95440 Bayreuth, Germany\\
$^6$ESRF, B.P. 220, 38043 Grenoble Cedex, France}
\date{\today}

\begin{abstract}
We study the change of the size and structure of freely suspended single lamella 
nanoparticles of polyethylene during thermal annealing  in aqueous solutions.
Using small-angle x-ray scattering and cryogenic transmission electron microscopy, 
it is shown that a doubling of the crystalline lamella sandwiched between two amorphous 
polymer layers is obtained by annealing the nanoparticles at $125\,^{\circ}\mathrm{C}$. 
This thickening of the crystalline lamella can be understood in terms 
of an unlooping of polymer chains within a single nanoparticle. In addition a variation 
of the annealing temperature from $90\,^{\circ}\mathrm{C}$ to $115\,^{\circ}\mathrm{C}$
demonstrates that the inverse of the crystalline lamellar thickness increases linearly 
with the annealing temperatures leading to a recrystallization line in a Gibbs-Thomson 
graph. Since the nanoparticles consist of about only eight polymer chains, they can 
be considered as a ideal candidates for the experimental realization of equilibrium 
polymer crystals.
\end{abstract}
\maketitle

\section{Introduction}

The crystallization of high molecular weight polymer chains such as polyethylene (PE) 
differs qualitatively from the crystallization of simple fluids in a number of important 
aspects such as entanglements. At a microscopic scale, entanglements arise from the fact that 
linear polymer chains are one-dimensionally connected objects which cannot cross each other. 
The resulting topological interaction strongly affects the crystallization process since it 
imposes constraints on the motion of the polymer chains. A perfect parallel alignment of all 
polymer chains in a crystalline state cannot be obtained starting from a melt because it would 
take too long time to disentangle them. By now it is well-established that the crystallization 
process leads to a two-phase structure consisting of platelike crystallites which are separated 
by amorphous regions. Within a crystallite, the polymer chains are aligned parallel to the main 
axis of the platelet, while the remaining entanglements are located in the amorphous regions. 
A given polymer chain may fold back into the same crystallite after a transit into the adjacent 
amorphous region because the chain length is considerably larger than the height of the platelike 
cystallites.

However, it is still under debate whether polymer crystallization is kinetically or 
thermodynamically controlled  (see, e.g., refs~
\cite{kell:94,hoff:97,kell:98,welc:01,muth:03,alle:05,somm:06,stro:06,stro:09,lan:10} 
and further references therein). The fact that the melting temperature and the crystallization 
temperature are different points to a decisive role of kinetics in polymer crystallization. 
However, computer simulations 
and theory \cite{welc:01,muth:03,somm:06,lari:05} have demonstrated the existence of 
equilibrium polymer crystals in the case that only a few polymer chains form a crystal. 
Polymer nanocrystals with small amorphous regions are ideal candidates for reaching 
thermodynamic equilibrium due the high mobility of the polymer repeat units.

Up to now, the overwhelming majority of studies have been done using bulk samples of PE. If 
the molecular weight is high enough, entanglements will play an important role for
 crystallization in these bulk samples.  Working with single crystals of the respective polymer 
has been a way around this problem and single crystals of PE have been studied since many years. 
\cite{kell:68}  Thus, crystals of PE with a thickness of about 10 - 15 nm and lateral dimensions 
of the order of 100 nm to a few $\mu$m can be generated by crystallization from a highly dilute 
solution. \cite{kwan:01,mago:03,mago:06} These crystals can be used to study the lamellar thickening 
of the semi-crystalline PE upon annealing. The morphological changes involved in these processes can 
be easily studied using atomic force microscopy. \cite{tian:01,mago:06} Thus, Tian and Loos observed 
that simultaneous thickening of single crystals of PE was accompanied by the formation of cavities 
within the crystal. \cite{tian:01} It is evident that research on single crystals allows us to monitor 
subtle morphological changes that take place upon annealing. However, in most investigations the PE 
crystals are lying on a solid substrate which may exert a decisive influence on the shape 
transformation (see the discussion of this point in ref~\cite{mago:06}).

Recently, the progress of catalytic polymerization has led to the generation of single crystals of 
PE with dimensions in the nanometric range. The catalytic polymerization technique yields stable 
aqueous suspensions of well-defined PE nanoparticles that can be studied by a wide variety of 
techniques in-situ. Thus, small platelike crystallites of PE have been prepared in this 
way. \cite{webe:07,chen:07}  Using a combination of cryogenic transmission electron microscopy 
(cryo-TEM) and small-angle x-ray scattering (SAXS), it has been demonstrated that these freely 
suspended PE platelets consist of a single crystalline lamella with a thickness of 6.3 nm. 
The overall thickness of 9 nm pointed to a small amorphous layer that could also be inferred precisely 
from the analysis of the SAXS-data. The facets of these crystals were clearly visible in micrographs 
taken by cryogenic transmission electron microscopy (cryo-TEM). Hence, the new method to create 
thin nanometric PE platelets provides the opportunity for new experiments on polymer crystallization 
using single PE crystals. \cite{chen:07}

Here we present the first study of the thickening of single lamellar PE nanoplatelets by thermal 
annealing. Earlier studies of this process involved either PE crystals supported by a solid 
substrate or bulk PE (see, e.g., 
refs~\cite{drey:70,barh:89,sadl:89,rast:97,xue:00,tian:01,tian:04,loos:06,naka:08} 
and references given therein). In particular, Tong et al. \cite{tong:09} presented the first 
systematic study of the lamellar thickening of PE nanocrystals on a solid surface by AFM. They 
showed that thermal annealing leads to considerable thickening of the crystals. Since the PE 
nanocrystals are freely suspended in an inert medium, 
any influence from solid substrates can be ruled out. Moreover, the small size allows us to monitor 
the overall shape and internal structure by a combination of cryo-TEM and SAXS. The present data 
can thus be compared to results obtained on solid substrates \cite{tian:01,tong:09} and on bulk samples.

\section{Experimental details and analysis of SAXS-data}

\subsection{Experimental}

The systems have been prepared by catalytic polymerization in aqueous solution as discussed 
in refs~\cite{baue:01,goet:06,yu:09}.
Our samples contain \mbox{1.6 wt \%} PE of molecular weight $3.5 \times 10^5$ g/mol and the
surfactant sodium dodecyl sulfate (SDS) to stabilize the PE particles against coagulation. 
The weight fractions of SDS are \mbox{0.87 wt \%} and \mbox{0.36 wt \%} for the two samples 
S87 and S36, respectively. For sample S87 the amount of SDS used during the 
synthesis of the PE particles has been increased as compared to our previous study\cite{webe:07}
in order to ensure colloidal stability even at rather high 
temperatures during annealing processes. Surface tension measurements have indicated that
virtually all surfactant molecules are adsorbed onto the strongly hydrophobic PE particles. 

Specimens for cryo-TEM have been prepared by vitrification of a thin liquid film of a PE 
dispersion supported by a copper grid in liquid ethane at its freezing point. Examinations 
were carried out at temperatures around  $-183\,^{\circ}\mathrm{C}$. Moreover, no staining 
agent has been used to enhance the contrast between the PE particles and the surrounding 
medium. All images have been recorded digitally by a bottom-mounted CCD camera system and 
processed with a digital imaging processing system.\cite{webe:07}

The SAXS measurements have been performed using either a custom-built Kratky compact camera or 
synchrotron radiation using the beamline ID02 at the ESRF. The scattering intensities of 
both the empty capillary and the solvent have been subtracted from the scattering intensities 
presented in the next section. As in our previous study a contrast variation between the PE 
particles and the solvent has been used by adding various amounts of glucose to the solutions. 
In this way the individual contributions of the amorphous and crystalline regions of the PE 
particles to the scattering intensity become available. This allows for consistency checks 
of the theoretical modelling.

\subsection{Scattering intensity}

SAXS determines the scattering intensity $I(q,\rho)$ as a function of the magnitude of the 
scattering vector $q$ and the PE particle number density $\rho$. For a system consisting of 
monodisperse PE particles the scattering intensity can be written as 
\begin{eqnarray} \label{eq1}
I(q,\rho)&=&\rho S(q,\rho) I_0(q) + \rho I_f(q)
\end{eqnarray}
where $I_0(q)$ describes how the scattering intensity is modulated by interference effects 
between radiation scattered by different parts of the same PE particle. The structure factor 
$S(q,\rho)$ is related to mutual interactions between different PE particles. Therefore it is 
dependent on the degree of order of the PE particles in the samples. For noninteracting particles 
the structure factor is unity. Moreover, the contribution to the scattering intensity due to
concentrations fluctuations of PE chains in the amorphous regions $I_f(q)$ becomes only important 
for large scattering vectors.

\begin{figure}[t!]
\begin{center}
\includegraphics[width=7cm,clip]{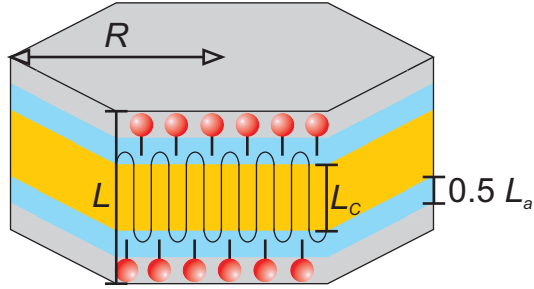}
\caption{Side view of a platelet of pseudo radius $R$ and total height $L$. 
The platelet consists of a crystalline lamella of height $L_c$ sandwiched between two 
amorphous sheets. These sheets include the hydrocarbon SDS tails with thicknesses 
$L_a/2$ and two layers of SDS head groups (marked in red).}
\label{fig1}
\end{center}
\end{figure}

The scattering intensity $I_0(q)$ of a single PE particle is described in terms of 
\begin{eqnarray} \label{eq2}
I_0(q)&=&\int\limits_0^1 d \alpha\, F^2(q,\alpha)
\end{eqnarray}
with
\begin{eqnarray} \label{eq3}
\lefteqn{F(q,\alpha)=}\nonumber
\\&&2\pi R \frac{J_1(qR\sqrt{1-\alpha^2})}{q\sqrt{1-\alpha^2}}
\times\left(\Delta b_{SDS} \frac{\sin (q\alpha L/2)}{q\alpha/2}\right.\nonumber
\\&& + \left. \Delta b_a \frac{\sin (q\alpha (L_a+L_c)/2)}{q\alpha/2}
     + \Delta b_c  \frac{\sin (q\alpha L_c/2)}{q\alpha/2}\right)
\end{eqnarray}
Here $R$ is the radius of a circular platelet consisting of a crystalline layer of height 
$L_c$ sandwiched between layers of amorphous PE and SDS as is shown in Figure \ref{fig1}. 
As already discussed in previous work, the faceted nanocrystals can be treated in good 
approximation as circular platelets for the SAXS analysis.\cite{webe:07} 
The height of the amorphous PE regions including the hydrocarbon SDS tails is given by 
$L_a/2$, while $L$ denotes the total height of the platelet (see Figure \ref{fig1}). The 
electron contrasts are defined as $\Delta b_{SDS}=b_{SDS}-b_s$, $\Delta b_a=b_a-b_{SDS}$, 
and $\Delta b_c=b_c-b_a$, where \mbox{$b_a = 302$ nm$^{-3}$}, \mbox{$b_c = 339$ nm$^{-3}$},
\mbox{$b_{SDS} = 396$ nm$^{-3}$}, and $b_s$ are the electron densities of the amorphous 
PE layers including the hydrocarbon SDS tails, the crystalline PE layer, the SDS headgroup 
layer, and the solvent, respectively. The value of the electron density of the solvent 
$b_s$ is determined by the concentrations of water and added contrast agent.
Moreover, $J_1(x)$ in eq \ref{eq3} denotes the cylindrical Bessel function of first order. 
The effect of size polydispersity of the PE particles is taken into account by an appropriate 
average using a distribution function characterizing the degree of polydispersity. 
%
%

We have used an integral equation theory \cite{harn:09} in order to calculate structure 
factors which characterize intermolecular correlations between different PE particles. This 
theoretical approach has been successfully applied to various suspensions consisting of 
platelike particles. \cite{harn:01,rose:06,harn:07}

Finally, the contribution of concentration fluctuations of the PE chains in the amorphous
phase reads 
\begin{eqnarray} \label{eq4}
I_f(q)&=&\frac{I_0^{(0)}}{1+(q\xi)^2}
\end{eqnarray}
where $\xi$ is the correlation length and $I_0^{(0)}$ determines the contribution 
at vanishing scattering vector.

\section{Results and Discussion}

\subsection{Increase of lamellar thickness}

\begin{figure}[t!]
\begin{center}
\hspace*{1.0cm}
\includegraphics[width=7cm,clip]{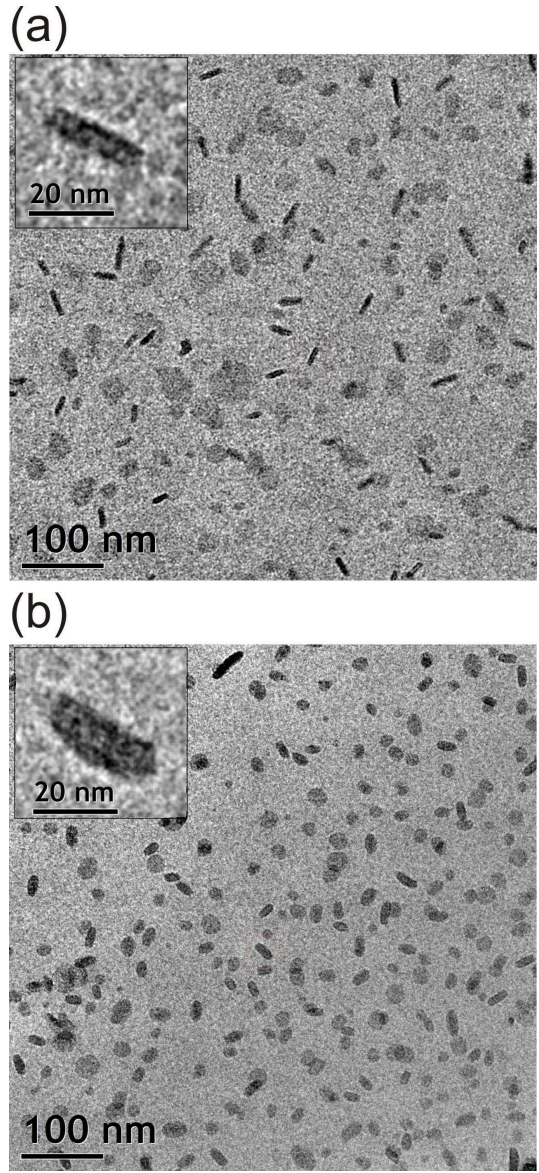}
\caption{Cryo-TEM micrographs of polyethylene nanoplatelets S87 in aqueous solution. 
The gray background is the low-density amorphous ice. In panel (a) the original 
sample was maintained at $25\,^{\circ}\mathrm{C}$ after catalytic polymerization 
at $15\,^{\circ}\mathrm{C}$, while in panel (b) the same sample was subject to 
an annealing process at $125\,^{\circ}\mathrm{C}$ for about 20 minutes. 
Thereafter, the sample was cooled down to $25\,^{\circ}\mathrm{C}$. The insets displays 
single particles in stronger magnification. Here the platelets are oriented with their 
normal parallel to the plane of the figure.}
\label{fig2}
\end{center}
\end{figure}

We first discuss the modification of the PE particles of sample S87 due to thermal 
annealing. A 5 ml glass bottle has been filled with about 2 ml of the original PE 
suspension. Thereafter, it has been kept in a metal vessel located on a heating plate 
at $125\,^{\circ}\mathrm{C}$ for 20 minutes. Finally, the sample has been 
cooled down to $25\,^{\circ}\mathrm{C}$. Evaporation of water could not be avoided 
completely leading to \mbox{1.74 wt \%} PE and \mbox{0.90 wt \%} SDS after annealing 
as compared to \mbox{1.70 wt \%} PE and \mbox{0.87 wt \%} SDS before annealing. 
Figures \ref{fig2} (a) and (b) show typical cryo-TEM micrographs of the suspension 
before and after annealing, respectively. The PE particles are platelets with rather 
narrow size distribution. The hexagonal facetting of the nanocrystals is clearly 
visible.\cite{webe:07} Evidently, the platelets do not form aggregates 
due to the added SDS molecules. Those PE particles that appear to be very close 
to each other are located along the optical path (perpendicular to the plane of the 
figures) but at different depths inside the suspension. Moreover, it is worthwhile 
to mention that the different gray scales for different platelets are also related 
to different angles between the normal of the platelets and the direction of the 
electron beam. The length of the optical path through a platelet with its normal 
oriented perpendicular to the electron beam is larger than that through a platelet 
with its normal oriented parallel to the electron beam.
%
%

From the cryo-TEM images we have determined platelet radii of 
\mbox{$14 \pm 4$ nm} and \mbox{$9 \pm 2$ nm} before and after annealing, respectively.
The corresponding heights of the platelets have been estimated to be 
\mbox{$7 \pm 1$ nm} and \mbox{$13 \pm 2$ nm} before and after annealing. These values 
have been derived from the image analysis of 20 particles with their normal oriented 
perpendicular to the electron beam. On the basis of this analysis one may conclude 
that a pronounced change of the shape of the PE nanoplatelets occurred during the 
annealing process (see also the insets of Figure \ref{fig2}). However, one has to take 
into account that it is not possible to 
detect amorphous PE with the help of the cryo-TEM micrographs shown in Figures 
\ref{fig2} (a) and (b) because the electron density of amorphous PE is very similar 
to the one of the surrounding low density amorphous ice. \cite{webe:07} However, the 
images shown in Figure \ref{fig2} demonstrate clearly that the process of annealing did 
not lead to formation of cavities, disintegration, or any other gross morphological 
change of the platelets. In order to elucidate the shape and the structure of the PE 
particles in more detail, the systems have been investigated using SAXS. 

\begin{figure}[t!]
\begin{center}
\includegraphics[width=7.5cm,clip]{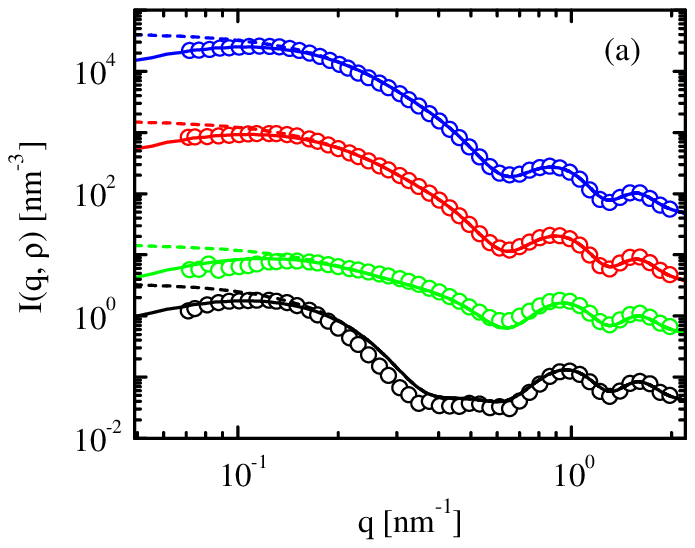}\\[15pt]
\includegraphics[width=7.5cm,clip]{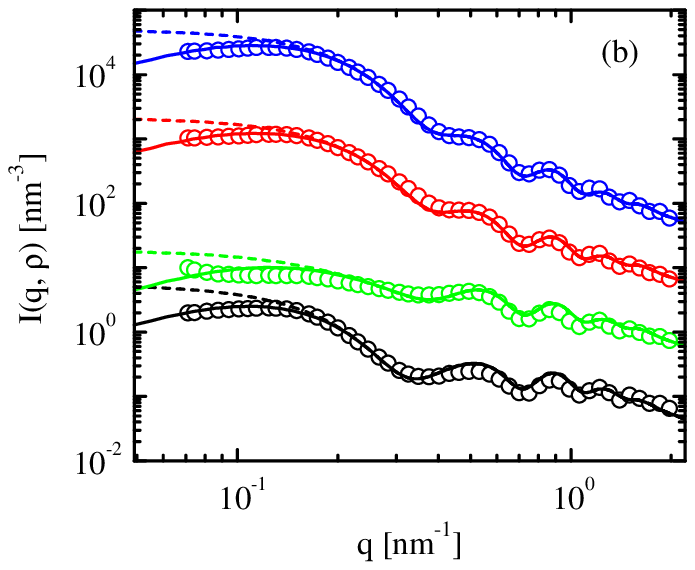}
\caption{Measured scattering intensity $I(q,\rho)$ of polyethylene nanoplatelets 
of sample S87 before (a) and after (b) annealing at $125\,^{\circ}\mathrm{C}$ as a 
function of the magnitude of the scattering vector q (symbols). The volume fraction of 
added glucose acting as contrast agent increases from bottom to top 
(0, 4.5, 10.0, 14.3 vol.~\%) while the volume fraction of the nanoplatelets decreases 
(2.6, 2.5, 2.3, 2.2 vol.~\% ). 
As a result the electron density of the solvent increases from bottom to top 
according to \mbox{$b_s = 333, 342, 352, 359$ nm$^{-3}$}. The 
three lowermost scattering intensities are shifted down by a factor of 10, 10$^2$, 10$^3$, 
respectively. The dashed lines represent the result of the modeling of the SAXS data 
assuming a dispersion of noninteracting polydisperse platelets. The solid lines 
represent the scattering intensity calculated for a dispersion of interacting  polydisperse 
platelets. The differences between the dashed and solid lines reflect the repulsive 
interaction between the nanoplatelets.}
\label{fig3}
\end{center}
\end{figure}

Figures \ref{fig3} (a) and (b) display SAXS intensities of the PE nanoplatelets before 
and after annealing, respectively. These scattering intensities have been measured at 
different contrasts starting from a stock solution of the nanoplatelets dispersed in 
pure water. The different contrasts have been adjusted by adding different amounts of 
glucose. The volume fraction of added glucose increases from 0 (lower symbols in 
Figures \ref{fig3} (a) and (b)) to 0.14 (upper symbols in Figures \ref{fig3} (a) and 
(b)) while the corresponding volume fractions of the nanoparticles decrease upon 
increasing the amount of added glucose. For clarity, the scattering intensities related 
to different contrast have been shifted vertically in Figure \ref{fig3}. 

Figure \ref{fig3} demonstrates that annealing the sample leads to marked 
differences in the scattering intensities. Moreover, contrast variation furthermore enhances 
the difference between these states. Hence, the change of shape and internal structure of 
the platelets can be studied with high precision. The lines show the results obtained from 
eqs \ref{eq1} - \ref{eq4} for noninteracting particles (dashed lines), i.e., $S(q,\rho)=1$, 
and interacting particles (solid lines). For small magnitudes of the scattering vector 
$q$ the calculated scattering intensities for noninteracting particles (dashed lines) 
on the one hand, and the integral equation results for interacting particles 
(solid lines) as well as the experimental data (symbols) on the other hand deviate 
due to strong repulsive electrostatic interactions between the particles brought 
about by the adsorbed SDS molecules. The correlation length associated with the peak of 
the scattering intensities at \mbox{$q = 0.14$ nm$^{-1}$} is given by 
\mbox{$d = 2\pi/q$ = 45 nm} and is related to the average distance between two platelets 
(see also Figure \ref{fig2}). Both this correlation length and the isothermal compressibility 
which is related to $I(0,\rho)$ \cite{harn:09} are the same before and after annealing. 
Hence, the intermolecular pair correlations and the equation of state of the suspension are
not influenced by the annealing process. This corroborates our finding that the number 
of platelets before and after annealing is equal (see Figure \ref{fig2}).
%
%

The overall dimensions following from the theoretical description of the scattering 
intensities are the average radius \mbox{$R = 10 \pm 3$ nm} and \mbox{$R = 7.5 \pm 3$ nm} 
as well as the thickness of the crystalline layer \mbox{$L_c = 6.5 \pm 1$ nm} and 
\mbox{$L_c = 13 \pm 1$ nm} before and after annealing, respectively. The thickness 
of the amorphous layer slightly increases from \mbox{$L_a = 3.1 \pm 0.8$ nm} to 
\mbox{$L_a = 3.8 \pm 1.0$ nm} during the annealing process. The change of the shape of 
the platelets can be directly seen from the shift of the location of the side maxima 
of the scattering intensities to lower $q$ values after annealing. The model parameters 
characterizing the size of the platelets can be expressed in terms of dimensionless 
scaling variables $qR$, $qL$, $q(L_a+L_c)$, and $qL_c$ according to equation \ref{eq3}. 
Hence a variation of the location of the $q$ values of the side maxima of the scattering 
intensity implies a change of these model parameters. The pronounced suppression 
of the scattering intensities for the samples with 4.5 vol.~\% added glucose is due to the 
fact that the electron density of the solvent, i.e., water and added glucose, is similar 
to that of the crystalline layer. Therefore, there is only a minor contribution of the 
crystalline layer to the scattering intensities in this case.
From the model parameters and by taking into account the densities of the amorphous and 
crystalline phase reported in the literature \cite{geil:63}, we have calculated that about 
$2\times 10^5$ CH$_2$ groups are forming each nanoparticle. Given the molecular weight 
$3.5 \times 10^5$ g/mol this corresponds to approximately 8 polyethylene chains by particle. 
Furthermore, we emphasize that we haven't found alternative models which lead to agreement 
with the experimental data shown in Figure \ref{fig3}.

\begin{figure}[h!]
\begin{center}
\includegraphics[width=7cm,clip]{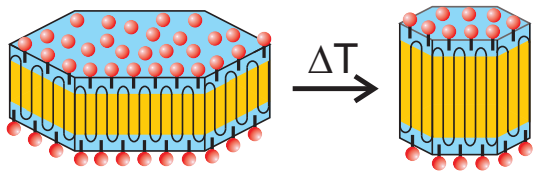}\\
\caption{Illustration of the mechanism leading to a doubling of a crystalline 
lamella after annealing. The figure shows schematically a side view of platelets 
consisting of an inner crystalline layer sandwiched between two amorphous layers.
The increase of the height of the crystalline lamella during annealing is due to 
an unlooping within a single lamella. \cite{drey:70,barh:89}}
\label{fig4}
\end{center}
\end{figure}

The analysis of the SAXS data and the TEM micrographs leads to the conclusion that annealing 
the sample S87 at $125\,^{\circ}\mathrm{C}$ for about 20 
minutes leads to a doubling of the thickness of the crystalline lamella from 
\mbox{$L_c = 6.5$ nm} to \mbox{$L_c = 13$ nm}. We have checked that longer annealing 
times (up to 60 minutes) do not lead to further increase of $L_c$. The thickening of the 
crystalline lamella during annealing can be understood in terms of an unlooping within a 
single lamella \cite{drey:70,barh:89} as is illustrated in Figure \ref{fig4}. With increasing 
temperature the chain mobility increases leading to cooperative motion of repeat units
parallel to the main axis of a crystalline platelet (see, e.g., 
refs~\cite{wang:09,rast:09} and references therein). In doing so the chains can 
partly unfold and reduce the amount of amorphous loops. As a result the height of the 
platelets increases while their radius decreases. 
%
%

It must be kept in mind that the number of particles inferred from SAXS is 
the same before and after annealing.  This is in agreement with the cryo-TEM micrographs. 
Within the present investigation this is an important observation because the alternative 
scenario, namely the stacking of adjacent lamella \cite{rast:97,xue:00} can be ruled out 
definitely. In this model the thickening is due to a fusion of two crystalline layers. 
Such a model may be invoked for solution crystallized films involving a semicrystalline 
polymer gel in which the crystalline domains are connected by amorphous material. For the 
present system a stacking of lamellae would lead to the reduction of the number of particle 
number by a factor two after annealing which is not observed.

For comparison we note that a simple theoretical model \cite{muth:03,somm:06}
for the equilibrium shape of a cylindrical polymer crystal yields 
\begin{eqnarray} \label{eq4a}
\frac{L_c}{R}=2 \frac{\sigma_{eff}}{\sigma_c}\,,
\end{eqnarray}
where $\sigma_c$ is the solvent-crystalline layer interfacial tension and 
$\sigma_{eff}$ is the effective solvent-amorphous layer interfacial tension 
that includes entropic and bending contributions \cite{harn:99a,harn:99b} of 
the PE chains. The temperature dependence of $\sigma_{eff}$ is discussed in Ref.~\cite{somm:06} 
on the basis of a statistical model that takes into account loops and tails of polymer 
chains in the amorphous regions. Depending on the model parameters $\sigma_{eff}$ decreases or 
increases upon increasing the temperature (see Figure 7 in Ref.~\cite{somm:06}). Equation 
\ref{eq4a} follows from a minimization of the free energy 
\begin{eqnarray} \label{eq4b}
F=-\pi R^2 L_c \epsilon_c + 2 \pi R L_c \sigma_c + 2 \pi R^2 \sigma_{eff}
\end{eqnarray}
with respect to $L_c$ under the constraint that $R^2 L_c=$ const. Here 
$\epsilon_c$ is the energy gain per volume of the bulk crystalline phase as 
compared to the bulk amorphous phase. After annealing, the shape of the 
PE nanocrystals under consideration is characterized by the ratio 
$L_c/R \approx 2$ implying that $\sigma_{eff}/\sigma_c \approx 1$ if the 
nanocrystals are thermodynamically stable. In view of the fact that no direct 
experimental values for the interfacial tensions are available, we note that 
$\sigma_{eff}/\sigma_c \approx 2$ for polymer crystal melt interfaces according 
to a very recent computer simulation study. \cite{miln:10} The presence of the 
SDS molecules and the solvent water will lead to a ratio 
$\sigma_{eff}/\sigma_c < 2$, but at the moment it is not possible to decide 
whether the PE nanocrystals after annealing exhibit their equilibrium shape 
or not. Nevertheless, the size ratio $L_c/R \approx 2$ of the PE nanocrystals 
is remarkably large as compared to the corresponding ratios of the well known 
PE crystals with a thickness of about 10 - 15 nm and lateral dimensions 
of the order of 100 nm to a few $\mu$m. Moreover, the scattering intensities shown 
in Figure \ref{fig3} (b) can be modelled using a size polydispersity with a constant 
size ratio $L_c/R$ according to equation \ref{eq4a}. Based on these results we 
see avenues for future research devoted to the experimental realization of equilibrium 
polymer nanocrystals.

\subsection{Variation of annealing temperature}

%
\begin{figure}[ht!]
\begin{center}
\includegraphics[width=7.5cm,clip]{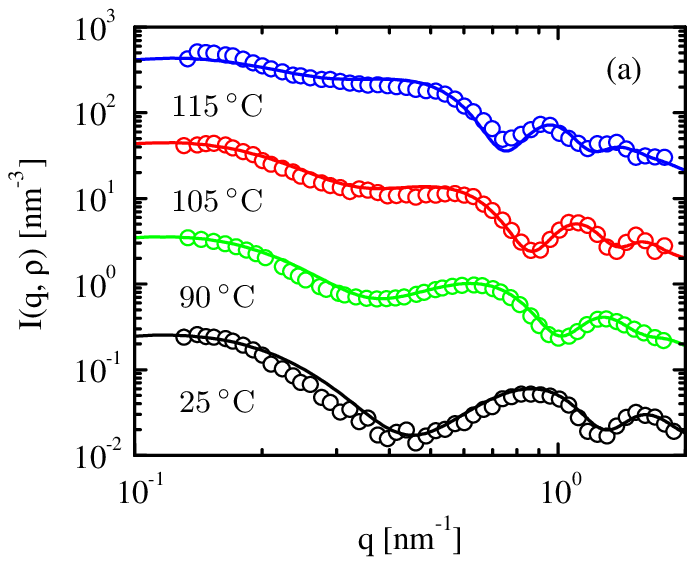}\\[15pt]
\includegraphics[width=7.5cm,clip]{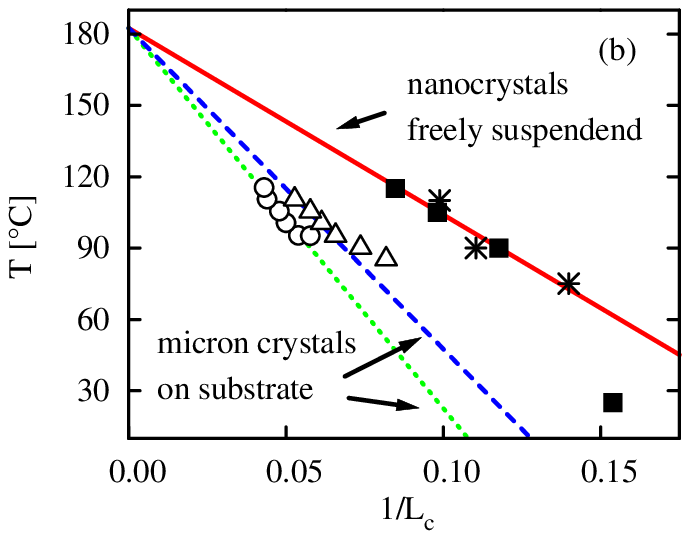}
\caption{(a) SAXS intensities of polyethylene nanoplatelets of sample S36
dispersed in pure water at $25\,^{\circ}\mathrm{C}$ together with the measured 
scattering intensities after annealing at $90\,^{\circ}\mathrm{C}$,
$105\,^{\circ}\mathrm{C}$, and $115\,^{\circ}\mathrm{C}$ (from bottom to top). 
For clarity, the three lowermost scattering intensities are shifted up by a factor 
of 10, 10$^2$, 10$^3$, respectively. The differences between the lower data set 
and the one shown in Figure \ref{fig3} (a) are due to the different amounts of 
added SDS in samples S36 and S87. The solid lines display the calculated 
results for interacting platelets. In (b) it is shown that the inverse of the 
crystalline lamellar thickness $1/L_c$ increases linearly with the annealing 
temperatures (upper three squares) leading to a recrystallization line (solid line).
The lower square corresponds to the original sample studied at $T=25\,^{\circ}\mathrm{C}$.
For comparison the open symbols display the thickness evolution during annealing of 
solution-grown micron crystals deposited on a solid substrate. \cite{tian:01}
The lowermost triangle and circle correspond to the micron crystals originally 
crystallized at $T=85\,^{\circ}\mathrm{C}$ and $T=95\,^{\circ}\mathrm{C}$, respectively. 
The stars mark the data taken from Figure 4 of ref. \cite{tong:09}.}
\label{fig5}
\end{center}
\end{figure}
%
The results of the previous subsection show that thermal annealing at the high 
temperature $125\,^{\circ}\mathrm{C}$ leads to a strong increase of the height of the 
crystalline lamella of the order of two. We now study possible changes of the PE 
platelets after annealing at lower temperatures using the sample S36 which contains 
a smaller amount of added SDS as compared to the sample S87 studied in the last subsection. 
Figure \ref{fig5} (a) displays SAXS intensities of the original sample together with 
the measured scattering intensities after annealing for about 20 minutes at 
$90\,^{\circ}\mathrm{C}$, $105\,^{\circ}\mathrm{C}$, and $115\,^{\circ}\mathrm{C}$ 
(from bottom to top). 

These scattering intensities have been obtained from a solution of PE platelets 
dispersed in pure water. For clarity, the three lowermost scattering intensities 
have been shifted vertically. The increase of the height of the nanoplatelets 
upon increasing the annealing temperature can be directly seen from the shift 
of the side maxima and minima to lower $q$-values. Longer annealing times (up to 
60 minutes) do not lead to further changes of the scattering intensities. We note 
that the differences between the measured scattering intensities of the original 
sample S36 (lower symbols in Figure \ref{fig5} (a)) and the original sample S87
(lower symbols in Figure \ref{fig3} (a)) are due to the different amounts of 
added SDS. However, the amount of added surfactant does not influence the size and 
structure of the PE nanoparticles.

The solid lines in Figure \ref{fig5} (a) show the calculated scattering 
intensities for interacting platelets using \mbox{$L_c = 6.5 \pm 1$ nm} for the 
original sample and \mbox{$L_c = 8.5 \pm 1.4$ nm}, \mbox{$L_c = 10.2 \pm 1.4$ nm},
\mbox{$L_c = 11.8 \pm 1.4$ nm} for the samples which have been annealed at 
$90\,^{\circ}\mathrm{C}$, $105\,^{\circ}\mathrm{C}$, and $115\,^{\circ}\mathrm{C}$,
respectively. Figure \ref{fig5} (b) displays a plot of the annealing temperature 
against the reciprocal thickness $L_c$ as suggested by the Gibbs-Thomson equation
\begin{eqnarray} \label{eq5}
T_c = T_c^{\infty}(1-\frac{2\sigma}{\Delta h L_c})
\end{eqnarray}
where $\Delta h$ is the heat of fusion, $\sigma$ the surface free energy of the lamellae, 
and $T_c^{\infty}$ the temperature limit referring to fully crystalline samples. 
The open symbols refer to the respective data taken from micron-sized PE crystals on a 
solid substrate. These data have been taken from the work of Tian and Loos. \cite{tian:01}
Evidently, the temperatures $T_c^{\infty}$ obtained by extrapolation of PE crystals of 
different sizes agree within prescribed limits of error. This is to be expected since 
the influence of both the SDS molecules in the case of the freely suspended nanocrystals 
and the solid substrate in the case of the micron crystals on the recrystallization 
process is vanishing in the limit of infinitely long polymer chains. Moreover, the data 
obtained by Tong et al.\cite{tong:09} on the thickening of PE-nanocrystals lying on a 
solid substrate (crosses in Figure \ref{fig5} (b)) fit very well on this line.

\begin{figure}[t!]
\begin{center}
\includegraphics[width=7.5cm,clip]{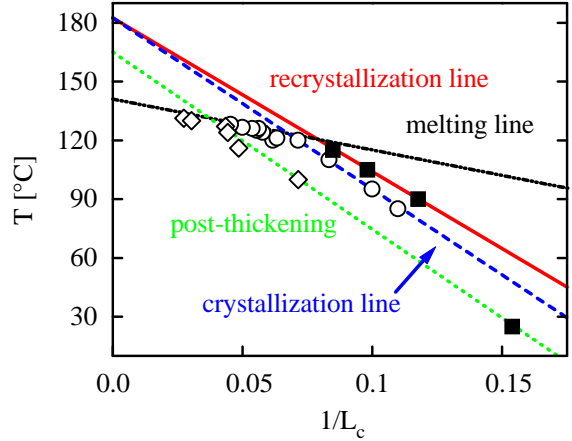}
\caption{Comparison of the recrystallization line (solid line) of freely suspended
polyethylene nanocrystals with the crystallization line (dashed line) of polyethylene 
crystallized from melt, while the dash-dotted line denotes the melting line. \cite{wund:77}
The open circles represent experimentally determined initial 
crystalline lamellar thicknesses of polyethylene crystallized from melt. \cite{barh:85} 
The open diamonds correspond to melt-crystallized polyethylene with already thickened 
crystalline lamella. \cite{voig:81} leading to the post-thickening line (dotted line). 
The freely suspended polyethylene nanoparticles went also through post-thickening after 
initial crystallization at room temperature (lowermost square).}
\label{fig6}
\end{center}
\end{figure}

It is also interesting to compare the lamella thickening of the PE nanocrystals with  
crystallization results obtained on PE bulk samples.
 Figure \ref{fig6} displays 
the recrystallization line (solid line) of the PE nanocrystals together with 
the crystallization line (dashed line) of PE crystallized from melt. \cite{barh:85} 
In the limit $L_c \to \infty$ the recrystallization line and the crystallization line 
lead to the same temperature $T_c^{\infty} \approx 182\,^{\circ}\mathrm{C}$. Note that 
here the initial thickness of the lamellae has been used for this plot 
(see the discussion of this point in ref. \cite{stro:05}). Together with the 
extrapolation shown in Figure \ref{fig5} this plot demonstrates that $T_c^{\infty}$ is a 
well-founded number describing the melting point of a fully crystalline PE-sample.
However, in case of bulk sample, the recrystallization process ends at the intersection of 
the recrystallization line with the melting line (dash-dotted line in Figure \ref{fig6}) 
which is given by $T = 141.1 - 259.7 / L_c$ according to ref~\cite{wund:77}.

For comparison the supercooling dependence of the crystalline lamellar thickness after 
post-crystallization reorganization is shown by the dotted line. After initial 
crystallization, the PE chains are known to reorganize toward the equilibrium state 
leading to an increase of the crystalline lamella thickness. The PE nanoparticles 
(lowermost square in Figure \ref{fig6}) went through post-thickening after initial 
crystallization. Extrapolation of these data according to eq. (\ref{eq5}) would lead to a 
considerably smaller value for $T_c^{\infty}$.

%
%

\section{Conclusion}

In conclusion, our findings elucidate the change of the size and the structure of 
individual polyethylene nanocrystals during annealing. The nanoparticles have been 
synthesized and stabilized by a nickel-catalyzed polymerization in aqueous solution. 
Hence freely suspended polymer nanocrystals can be studied, while earlier studies 
of larger polyethylene crystals involved either the bulk polymer phase or a 
supporting solid substrate. The combination of small-angle x-ray scattering and 
cryogenic transmission electron microscopy demonstrates that the thickening of the 
crystalline lamella can be controlled by varying the annealing temperature. The 
resulting recrystallization line defines a linear relationship between the annealing 
temperature and the reciprocal of the crystalline lamellar thickness. The
recrystallization lines for both the present PE nanocrystals and for micron-sized 
PE crystals (see Figure \ref{fig5}) lead to the same temperature 
$T_c^{\infty} \approx 182\,^{\circ}\mathrm{C}$ as the crystallization line for 
for bulk samples (see Figure \ref{fig6}). All data obtained herein point to a thermodynamic 
control of the thickness of the lamellae. The finite thickness measured after annealing 
can be rationalized by a simple theoretical model \cite{muth:03,somm:06} for the 
equilibrium shape of a cylindrical polymer crystal.

\section{Acknowledgment}
This work was partly funded by the German Excellence Initiative (Q.T. and S.M.). 
M.B. acknowledges gratefully financial support by the DFG.

\end{document}